\documentclass[aps,prl,reprint,superscriptaddress]
{revtex4-1}

\usepackage{graphicx}
\usepackage{float}
\usepackage{gensymb}
\usepackage[mathscr]{euscript}
\usepackage{amsmath}

\begin{document}

\title{Fluctuation spectroscopy as a probe of granular superconducting diamond films}

\author{G. M. Klemencic}

\affiliation{School of Physics and Astronomy, Cardiff University, Queen's Buildings, The Parade, Cardiff, CF24 3AA, UK}

\author{J. M. Fellows}
\affiliation{School of Physics, HH Wills Physics Laboratory, University of Bristol, Tyndall Avenue, Bristol, BS8 1TL, UK}

\author{J. M. Werrell}
\affiliation{School of Physics and Astronomy, Cardiff University, Queen's Buildings, The Parade, Cardiff, CF24 3AA, UK}

\author{S. Mandal}
\affiliation{School of Physics and Astronomy, Cardiff University, Queen's Buildings, The Parade, Cardiff, CF24 3AA, UK}

\author{S. R. Giblin}
\affiliation{School of Physics and Astronomy, Cardiff University, Queen's Buildings, The Parade, Cardiff, CF24 3AA, UK}

\author{R. A. Smith}
\affiliation{School of Physics and Astronomy, University of Birmingham, Edgbaston, Birmingham B15 2TT, UK}

\author{O. A. Williams}
\affiliation{School of Physics and Astronomy, Cardiff University, Queen's Buildings, The Parade, Cardiff, CF24 3AA, UK}


\date{\today}

\begin{abstract}
We present resistance versus temperature data for a series of boron-doped nanocrystalline diamond films whose grain size is varied by changing the film thickness. Upon extracting the fluctuation conductivity near to the critical temperature we observe three distinct scaling regions -- 3D intragrain,  quasi-0D, and 3D intergrain --  in confirmation of the prediction of Lerner, Varlamov and Vinokur. The location of the dimensional crossovers between these scaling regions allows us to determine the tunnelling energy and the Thouless energy for each film. This is a demonstration of the use of \emph{fluctuation spectroscopy} to determine the properties of a superconducting granular system.
\end{abstract}
\maketitle

Tunable granular materials offer rich physical systems with which to study the interplay between electron correlations and the mesoscopic effects of disorder. The occurrence of the metal-insulator and superconductor-insulator transitions appear to be strongly linked to granularity, be it structural or pertaining to variations of the order parameter~\cite{gantmakher2010superconductor, sacepe2008disorder, jaeger1989onset}.  There are also clear theoretical predictions for the signature of granularity in the transport properties of disordered superconductors close to the superconducting transition~\cite{beloborodov2007granular, lerner2008fluctuation, deutscher1974superconducting, imry1981destruction}. Boron doped nano-crystalline diamond (BNCD) provides a suitable tuneable material in which to explore these theoretical predictions. Superconductivity was first observed in high-pressure, high-temperature fabricated boron doped diamond in 2004~\cite{ekimov2004superconductivity}. The phenomenon was quickly demonstrated in both single-crystalline~\cite{bustarret2004dependence} and polycrystalline~\cite{takano2004superconductivity} diamond synthesised by chemical vapour deposition (CVD). While superconductivity in doped semiconductor materials~\cite{blase2009superconducting, bustarret2015superconductivity} is an active area of research, and although nanocrystalline diamond retains many of the desirable mechanical properties of single-crystalline material~\cite{williams2010high}, a sometimes overlooked property in the study of superconductivity in polycrystalline boron-doped diamond is the physical granularity itself. 

A clear experimental signature of superconducting granular systems, as pointed out by Lerner, Varlamov and Vinokur \cite{lerner2008fluctuation} (henceforth referred to as LVV), is that there are three distinct temperature regimes in the vicinity of the critical temperature ($T_c$), distinguished by the magnitude of the temperature-dependent Ginzburg-Landau coherence length. At temperatures immediately above $T_c$, short-lived Cooper pairs act as charge carriers, modifying the conductivity. The principal modification close to $T_c$ is the Aslamazov-Larkin pair contribution to the conductivity -- the so-called paraconductivity~\cite{larkin2005theory}. When the coherence length is much larger than the typical grain size, the granularity is not seen by a Cooper pair and the system behaves as a 3D superconductor with paraconductivity taking the well-known form $\propto\epsilon^{-1/2}$, where $\epsilon=(T-T_c)/T_c$ is the reduced temperature. When the coherence length is comparable to the typical grain size, each grain acts as its own 0D superconductor for which the paraconductivity is expected to be $\propto\epsilon^{-2}$. However, in addition, the intergrain transport requires two electrons to hop with a cooper pair lifetime $\propto1/\epsilon$ such that the conductivity in a quasi-0D (q0D) array of grains should in fact be $\propto\epsilon^{-3}$. Finally, when the coherence length is much smaller then the typical grain size, a Cooper pair moves through the grain as though it were a bulk 3D system so that the conductivity is again $\propto\epsilon^{-1/2}$ albeit reduced by renormalization of the scattering time due to scattering at grain boundaries~\footnote{In the original LVV paper this exponent is miscalculated as $-2$.}. Although these scaling laws are presented on physical grounds, they are calculated directly from the appropriate Aslamazov-Larkin diagram in the original LVV paper, accounting for intergrain tunnelling; hence the theory is both intuitive and rigorous.  Shifting perspective from length scales to reduced temperature scales, the LVV theory predicts a dimensional crossover from 3D to q0D behaviour at $\epsilon\sim\epsilon_\text{t}=\Gamma/k_\textsc{b}T_c$, where $\Gamma$ is the tunnelling energy, and a second crossover from q0D back to 3D at $\epsilon\sim\epsilon_\text{g}=E_\text{Th}/k_\textsc{b}T_c$, where $E_\text{Th}=\hbar\mathscr{D}/a^2$ is the Thouless energy, $\mathscr{D}$ is the intragrain diffusion constant, and $a$ is the typical grain size. To summarise, the LVV theory predicts a fluctuation paraconductivity, $\sigma_\text{fl}$, with scaling behaviour
\begin{equation}
\frac{{\rm d}\ln\sigma_\text{fl}}{{\rm d}\ln\epsilon} = 
\begin{cases}
-\tfrac{1}{2} \qquad{(3D)}\quad& \epsilon \lesssim \epsilon_\text{t} \\
-3 \qquad{(q0D)}\quad& \epsilon_\text{t} \lesssim \epsilon \lesssim \epsilon_\text{g} \\
-\tfrac{1}{2} \qquad{(3D)}\quad& \epsilon_\text{g} \lesssim \epsilon \lesssim 1\quad. \\
\end{cases}
\end{equation}

In this Letter, by measuring the resistive transition of BNCD films, we experimentally observe a clear signature of granularity due to different regimes of superconducting fluctuation contributions at $T \gtrsim T_c$. To this end, we present a detailed fluctuation spectroscopy analysis of transport measurements on a series of BNCD films with increasing grain sizes as controlled by film thickness. We observe the dimensional crossovers with $\epsilon$ scaling predicted by the LVV theory, extracting the Thouless energy, tunnelling energy, and the inter- and intragrain diffusion constants. Further to this observation, we anticipate that the considerable morphological control afforded by the tunable growth of superconducting BNCD~\cite{williams2011nanocrystalline} provides a rich physical system with which to further test these theoretical models. Dimensional crossovers in superconducting materials have been investigated in the past~\cite{xiang1993three, deutscher1977critical, makise2012fluctuation} but not used as a method of material characterisation. The close agreement between theory and experiment shown here allows the extraction of various physical characteristics of films; this simple method of determining physical parameters should be useful in selecting materials for device applications such as superconducting single photon detectors (SSPDs) and kinetic inductance detectors (KIDs) amongst others~\cite{natarajan2012superconducting, day2003broadband, shurakov2015superconducting}.

The exact structural granular morphology of a CVD-grown BNCD film is sensitive to a number of parameters. To efficiently grow a thin uniform film on a non-diamond substrate, the surface must first be populated with a high density of nucleation sites - typically small diamond particles - from which the film's composite grains may grow~\cite{williams2007enhanced}. The density and initial size of particles are significant factors in the grain sizes of the resultant film, which is formed by the competitive growth of the seed crystals in a plasma comprised of a dilute concentration of methane in hydrogen. A hydrogen rich environment preferentially etches non-diamond material; the grains grow in a columnar structure originating from the particular nucleation sites, and increase in size as the film grows thicker~\cite{willems2010granular}. By changing the ratio of methane and hydrogen in the plasma, larger or smaller grains can be grown, for a lower or higher methane percentage respectively~\cite{zhang2011role}. In addition to this, CVD allows control of the growth temperature and chamber pressure. All of these parameters, and the interplay between them, determine the granularity of the nanodiamond film.

While intrinsic diamond is a wide-bandgap insulator~\cite{celii1991diamond}, the incorporation of boron above a critical value, $n_c \sim 4.5 \times 10^{20}$~cm$^{-3}$, provides the necessary charge carriers for superconductivity~\cite{klein2007metal}. There is some evidence to suggest that boron is incorporated more efficiently in some growth directions than others~\cite{ushizawa1998boron}, and that an adjustment of the methane concentration in BNCD growth can affect the overall dopant concentration leading to the reporting of some contradictory results~\cite{willems2010granular, zhang2011role}. Given the tunability of CVD diamond growth, it is no surprise that there is a wide variation in the properties of polycrystalline superconducting diamond reported in the literature~\cite{bustarret2008superconducting, achatz2009low, dahlem2010spatially, gajewski2009electronic, klein2007metal, takano2004superconductivity, willems2010granular, zhang2011role}. It is exactly this tunability, however, that is a powerful tool in the well-controlled production of this granular superconducting material.

We grew a series of BNCD films on SiO$_2$ buffered (100) silicon wafers by microwave plasma assisted CVD~\cite{williams2008growth}. Prior to growth, each substrate was seeded by ultrasonic agitation in a monodisperse aqueous colloid of nanodiamond particles of approximately 5~nm diameter. This process has been shown to provide a nucleation site density in excess of 10$^{11}$~cm$^2$~\cite{williams2007enhanced}. The substrates were held at $\sim$720$\degree$C in a dilute gas mixture of methane and trimethylboron in hydrogen, with a 3\% methane concentration and a B/C ratio of 12800~ppm, and a chamber pressure and microwave power of 40~Torr and 3.5~kW respectively. The growth time was varied across the set such that the film thickness was in the range 35--564~nm; the B/C ratio was not changed between samples, thus we expect similar boron concentrations across the series~\cite{willems2010granular}. 

\begin{figure}
\includegraphics[width=8.6cm]{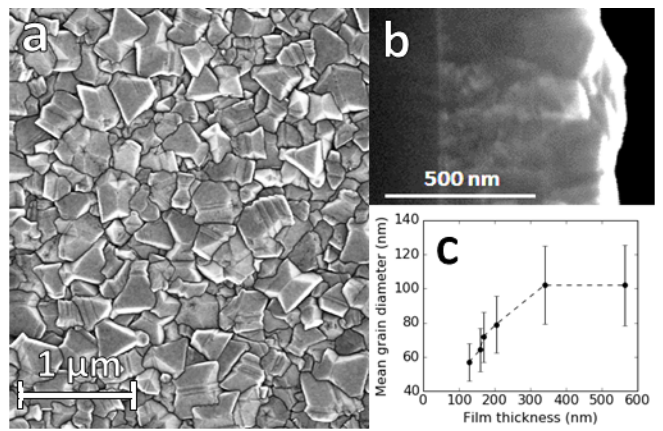}
\caption{\label{SEM}(a) Surface morphology of 564~nm thick nanocrystalline diamond film imaged by SEM. (b) The thickness was determined by imaging a cross section of the film. (c) Grain size analysis gave a measure of the mean grain diameter, shown as a function of film thickness.}
\end{figure}

\begin{figure}
\includegraphics[width=8.6cm]{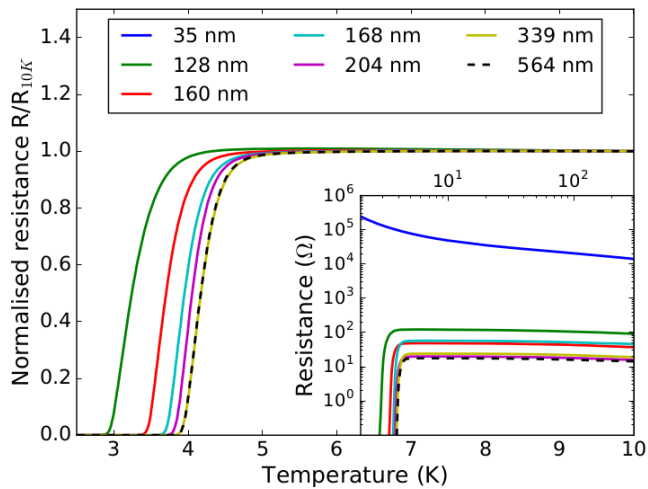}
\caption{\label{RT_300K_10K}Resistive transition of thickness-varied series of BNCD films normalised to $R_{10K}$. The thinnest film is omitted due to lack of observable superconductivity in this temperature range. Inset shows $R(T)$ from room temperature to 2~K. }
\end{figure}

Scanning electron microscope (SEM) images were used to quantify the granularity of the BNCD films. As an example, Fig.~\ref{SEM}(a) shows the surface morphology of a 564~nm thick film. The polycrystalline nature of the material is evident, with no single dominant growth direction observable. Fig.~\ref{SEM}(b) shows a cross sectional image of the same film used to determine the thickness; the columnar grain structure and surface roughness is also evident. The lateral surface grain size distribution determined from the SEM images was used to give a mean grain diameter as a function of film thickness, shown in Fig.~\ref{SEM}(c). It is clear that thicker films have a larger mean grain size, although detailed grain size analysis revealed that the 339~nm and 564~nm thick films have a similar grain size distribution at the surface. The temperature dependence of the sample resistances, shown in Fig.~\ref{RT_300K_10K}, were measured in the range 2-300~K using a Quantum Design Physical Property Measurement System. Silver paste contacts were made to the sample surface in a four-wire Van der Pauw configuration. The $R(T)$ curves in Fig.~\ref{RT_300K_10K} show an overall increase in $T_c$ as the film thickness is increased; no further change in $T_c$ is observed in thicker samples.

\begin{figure}
\includegraphics[width=8.6cm]{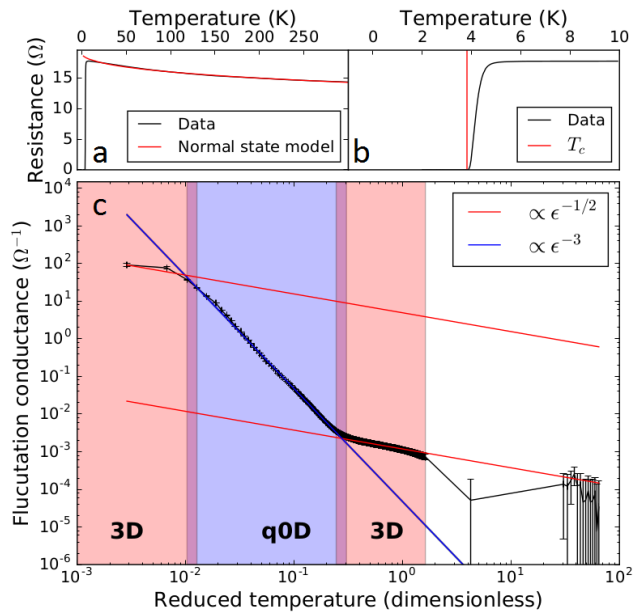}
\caption{\label{Analysis_example}Fluctuation spectroscopy of a 564~nm thick BNCD film. (a) A fit to the high temperature region reveals a $T^{0.5}$ dependence, and allows extraction of $G_\text{ns}$. (b) $T_c$ is defined as the point at which the conductance diverges, depicted in red. (c) Fluctuation conductance, $G_\text{fl} = G -G_\text{ns}$, as a function of reduced temperature, $\epsilon = (T-T_c)/T_c$.} 
\end{figure}

In order to extract the fluctuation contribution to the conductivity, we first subtract the normal-state temperature dependence. The high temperature conductivity in granular diamond has a pronounced $\sqrt{T}$ behaviour,  shown in Fig.~\ref{Analysis_example}(a), as might be expected for a disordered system with an appreciable electron-electron interaction~\cite{altshuler1985electron}. This has previously been observed in single crystal diamond samples~\cite{bousquet2016phase}. We therefore extract the normal state contribution by fitting the high temperature conductance data to $G_\text{ns}=a+b\sqrt{T}$.
The fluctuation conductance in the vicinity of the transition is therefore $G_\text{fl} = G- G_\text{ns}$. We determine the transition temperature, $T_c$, as the point at which $G_\text{fl}$ diverges, shown in Fig.~\ref{Analysis_example}(b) as a red vertical line on the $R(T)$ curve for a 564~nm thick film. This fluctuation analysis is in contrast to the typical experimental procedure of defining $T_c$ as the midpoint of the transition region, but is necessary if we are to correctly assign the reduced temperature. Fig.~\ref{Analysis_example}(c) shows the fluctuation conductance as a function of the reduced temperature, plotted on a logarithmic scale. The emergence of distinct regions of different scaling behaviour is immediately observable, supporting LVV-theory. This behaviour is observable in every film in the series; Fig.~\ref{Analysis_example}(c) is representative of the data obtained. Similar results of fluctuation spectroscopy for all films are shown in Fig. 1--6 in Supplementary Material.

From the distinct temperature regimes observable in the fluctuation conductivity, the Thouless energy, $E_\text{Th}$, and the tunnelling energy, $\Gamma$, can immediately be extracted. As the crossover is smooth, one cannot simply find a point at which it occurs. Instead we define a crossover region as that in which two scaling forms can be fit within 10\% of the experimental values. We then use the upper and lower boundaries to estimate the error on the crossover point. In Fig.~\ref{Analysis_example}(c), the crossover regions found in this manner are shown as the regions where the two shadings overlap. While there are few data points in the dimensional regime immediately following $T_C$, it is clear that there exists a region with a fluctuation conductance that is quantitatively different to the q0D regime that follows. We note that despite having the fewest points, the scaling behaviour in the bulk 3D regime follows Ginzburg-Landau theory near the transition temperature and as such is expected.

\begin{figure}
\includegraphics[width=8.6cm]{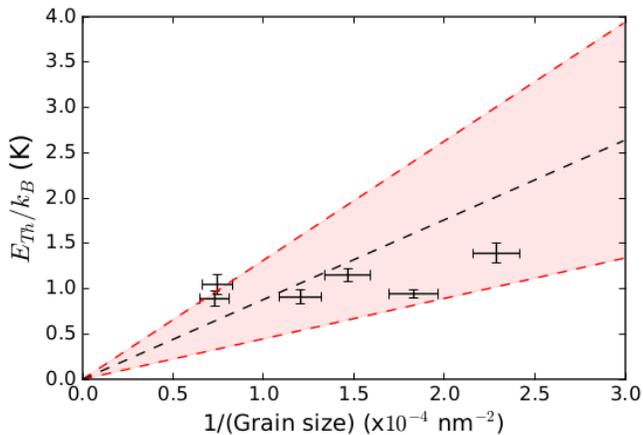}
\caption{\label{Diffusion_constant}Thouless energy, $E_{Th}$, as a function of $1/a^2$, where $a$ is the mean grain diameter. A linear fit is used to calculate the electron diffusion constant as $\mathscr{D} = $11.5$\pm$5.7 cm$^2$/s. The shaded area represents the error on the linear fit.}
\end{figure}

The tunnelling and Thouless energies were calculated based on these crossovers for each film. The tunnelling energy was extracted from the 3D to q0D crossover for each sample and found to be $\Gamma = 4.2\pm 2.0$~$\mu$eV; this is the energy associated with charge carriers crossing a grain boundary, and as such is not expected to change significantly from film to film. In contrast, by definition, the Thouless energy should be proportional to the inverse square of the mean grain size, with the constant of proportionality giving the intragrain diffusion constant. In Fig.~\ref{Diffusion_constant}, the Thouless energy is extracted from the q0D to 3D crossover for each film, and is plotted as a function of $a^{-2}$. The shaded region represents the error on the linear fit; it is important to note here that the main contribution to the uncertainty is the grain size distribution as quantified by SEM image analysis. This result serves as a consistency check of the underpinning physics; the extracted values of $E_\text{Th}$ fall within the the range given by the spread of grain sizes obtained from SEM analysis, thus demonstrating that our measurement is consistent with the theoretical prediction that describes them. The intragrain diffusion constant, $\mathscr{D}$, is calculated from the gradient of this plot as 11.5$\pm$5.7 cm$^2$/s. The bulk diffusion constant for a BNCD film grown under the same conditions was recently measured to be 0.6~cm$^2$/s~\cite{titova2016slow}; if we replace the Thouless energy with the tunnelling energy, we obtain the effective intergrain diffusion constant defined as $\mathscr{D}_\text{eff}=a^2\Gamma/\hbar$ \citep{beloborodov2007granular} which we find to be 0.54$\pm$0.36  cm$^2$/s, showing that this method of extracting physical parameters is consistent with other approaches.

To conclude, we have applied a theoretical model of the fluctuation paraconductivity in granular superconductors to a series of BNCD films where the grain size was varied by changing the film thickness. We demonstrate an excellent agreement between LVV theory and experimental measurements of $R(T)$ in BNCD films over a wide temperature range. The clear observation of dimensional crossover in the fluctuation region allows conclusions to be drawn about the physical influence of the structural granularity of the material. We present a simple graphical method of estimating the tunnelling energy and the Thouless energy, leading to an estimate of the inter- and intragrain electron diffusion constants respectively. Knowledge of these parameters is an important consideration in the design of superconducting electronic devices such as SSPDs or KIDs. We propose that the technique of fluctuation spectroscopy demonstrated here may prove to be a remarkably simple yet valuable tool for characterising the microscopic properties of BNCD and other granular superconducting materials.

\begin{acknowledgments}
The authors wish to thank A. Papageorgiou for help with image processing. We gratefully acknowledge support by the European Research Council under the EU Consolidator Grant `SUPERNEMS'.
\end{acknowledgments}


\bibliography{BDDGranularSuperconductivity}

\widetext
\clearpage

\begin{center}
\textbf{\large Supplemental Materials: \\ Fluctuation spectroscopy as a probe of granular superconducting diamond films}
\end{center}
\setcounter{equation}{0}
\setcounter{figure}{0}
\setcounter{table}{0}
\setcounter{page}{1}
\makeatletter
\renewcommand{\theequation}{S\arabic{equation}}
\renewcommand{\thefigure}{S\arabic{figure}}
\renewcommand{\bibnumfmt}[1]{[S#1]}
\renewcommand{\citenumfont}[1]{S#1}

\begin{figure}[!h]
\includegraphics[width=8.6cm]{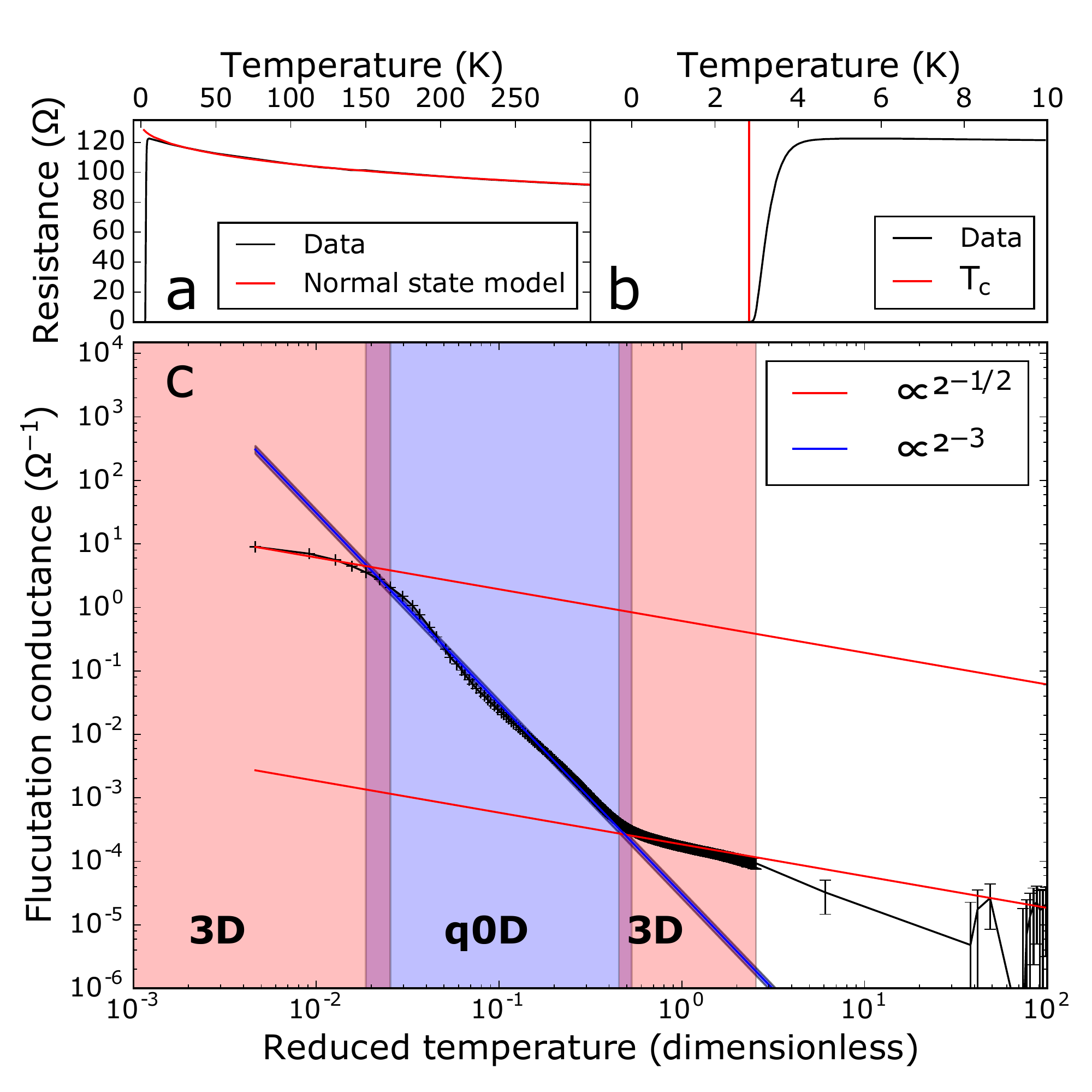}
\caption{\label{128nm}Fluctuation spectroscopy of a 128~nm thick BNCD film. (a) A fit to the high temperature region reveals a $T^{0.5}$ dependence, and allows extraction of $G_\text{ns}$. (b) $T_c$ is defined as the point at which the conductance diverges, depicted in red. (c) Fluctuation conductance, $G_\text{fl} = G -G_\text{ns}$, as a function of reduced temperature, $\epsilon = (T-T_c)/T_c$.}
\end{figure}

\begin{figure}
\includegraphics[width=8.6cm]{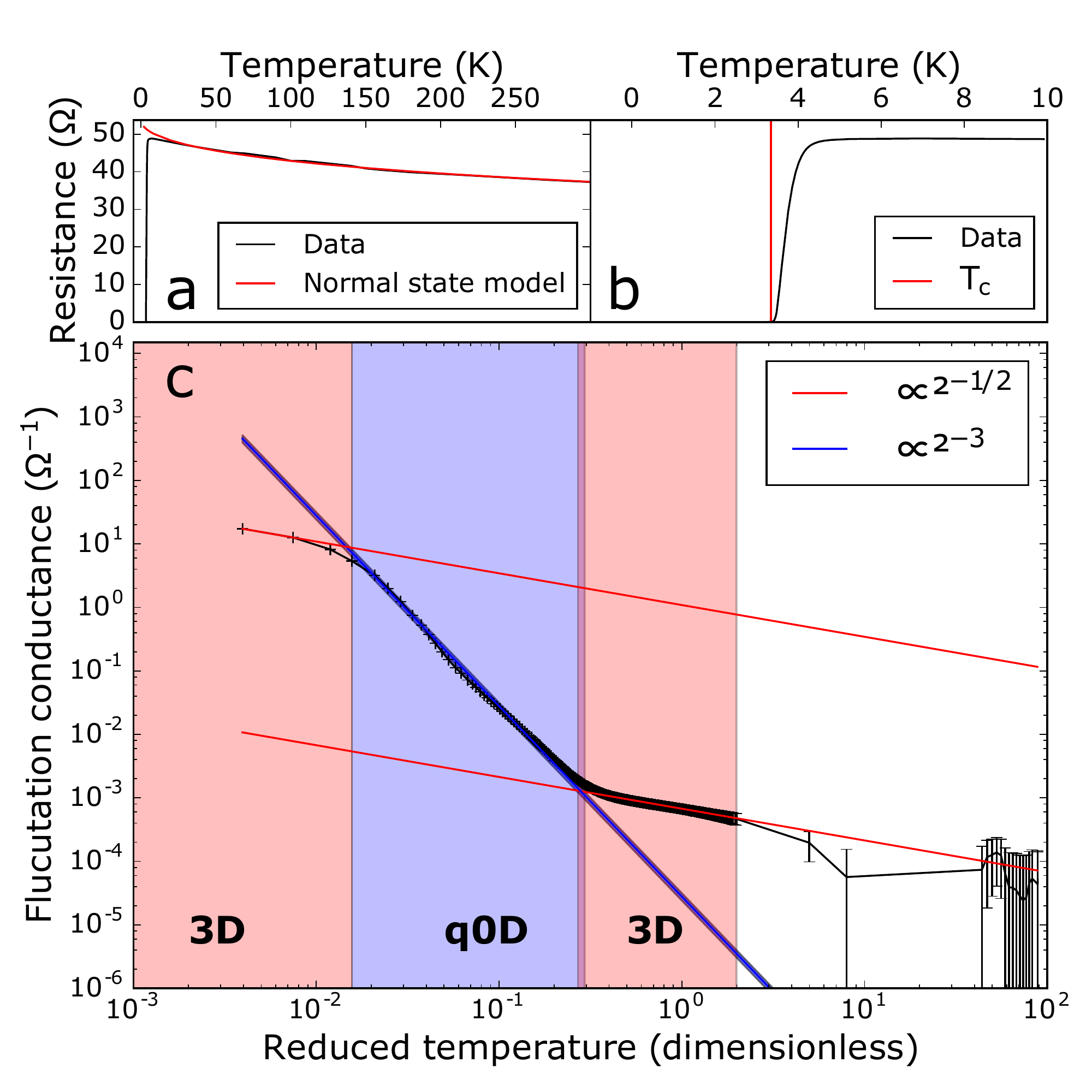}
\caption{\label{160nm}Fluctuation spectroscopy of a 160~nm thick BNCD film. (a) A fit to the high temperature region reveals a $T^{0.5}$ dependence, and allows extraction of $G_\text{ns}$. (b) $T_c$ is defined as the point at which the conductance diverges, depicted in red. (c) Fluctuation conductance, $G_\text{fl} = G -G_\text{ns}$, as a function of reduced temperature, $\epsilon = (T-T_c)/T_c$.}
\end{figure}

\begin{figure}
\includegraphics[width=8.6cm]{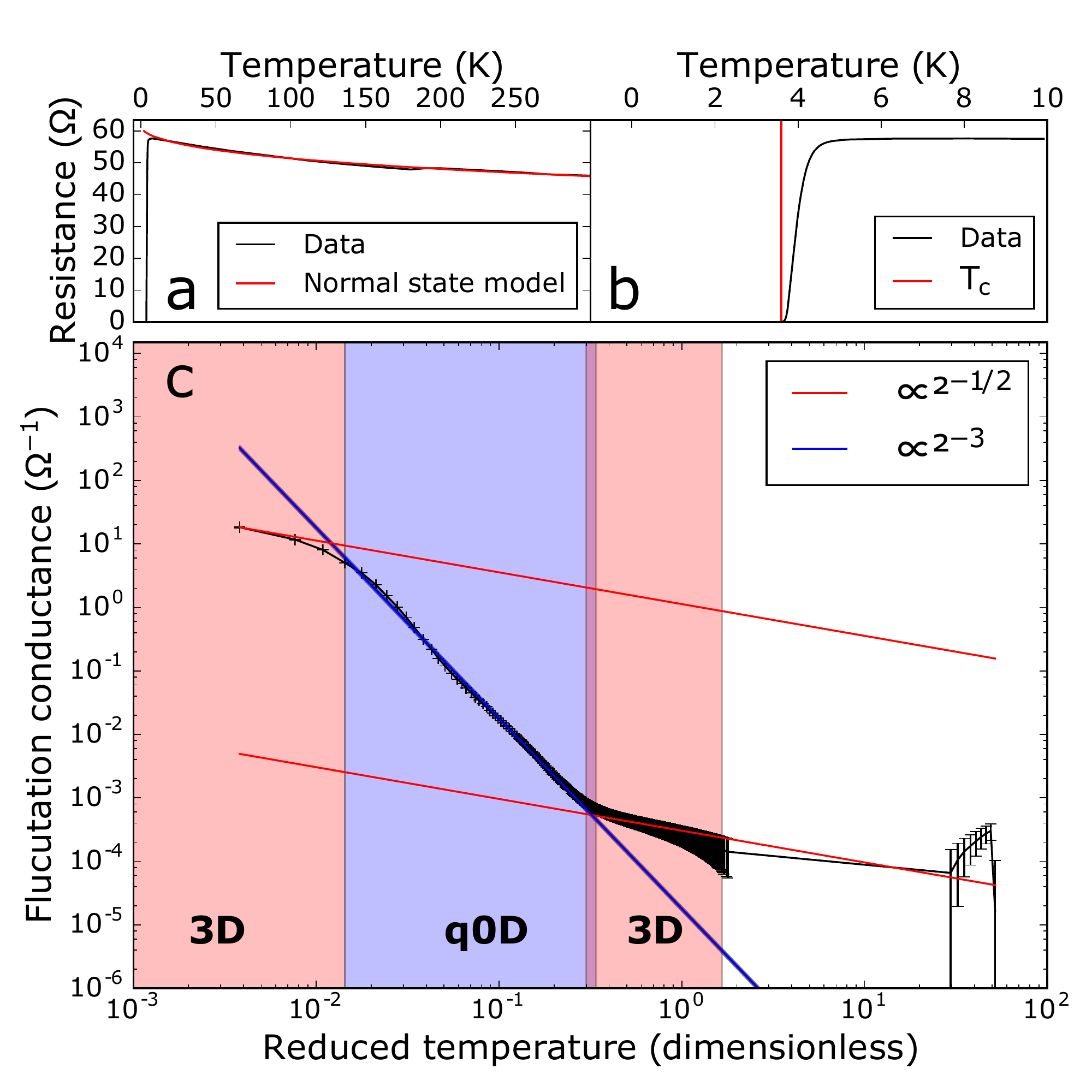}
\caption{\label{168nm}Fluctuation spectroscopy of a 168~nm thick BNCD film. (a) A fit to the high temperature region reveals a $T^{0.5}$ dependence, and allows extraction of $G_\text{ns}$. (b) $T_c$ is defined as the point at which the conductance diverges, depicted in red. (c) Fluctuation conductance, $G_\text{fl} = G -G_\text{ns}$, as a function of reduced temperature, $\epsilon = (T-T_c)/T_c$.}
\end{figure}

\begin{figure}
\includegraphics[width=8.6cm]{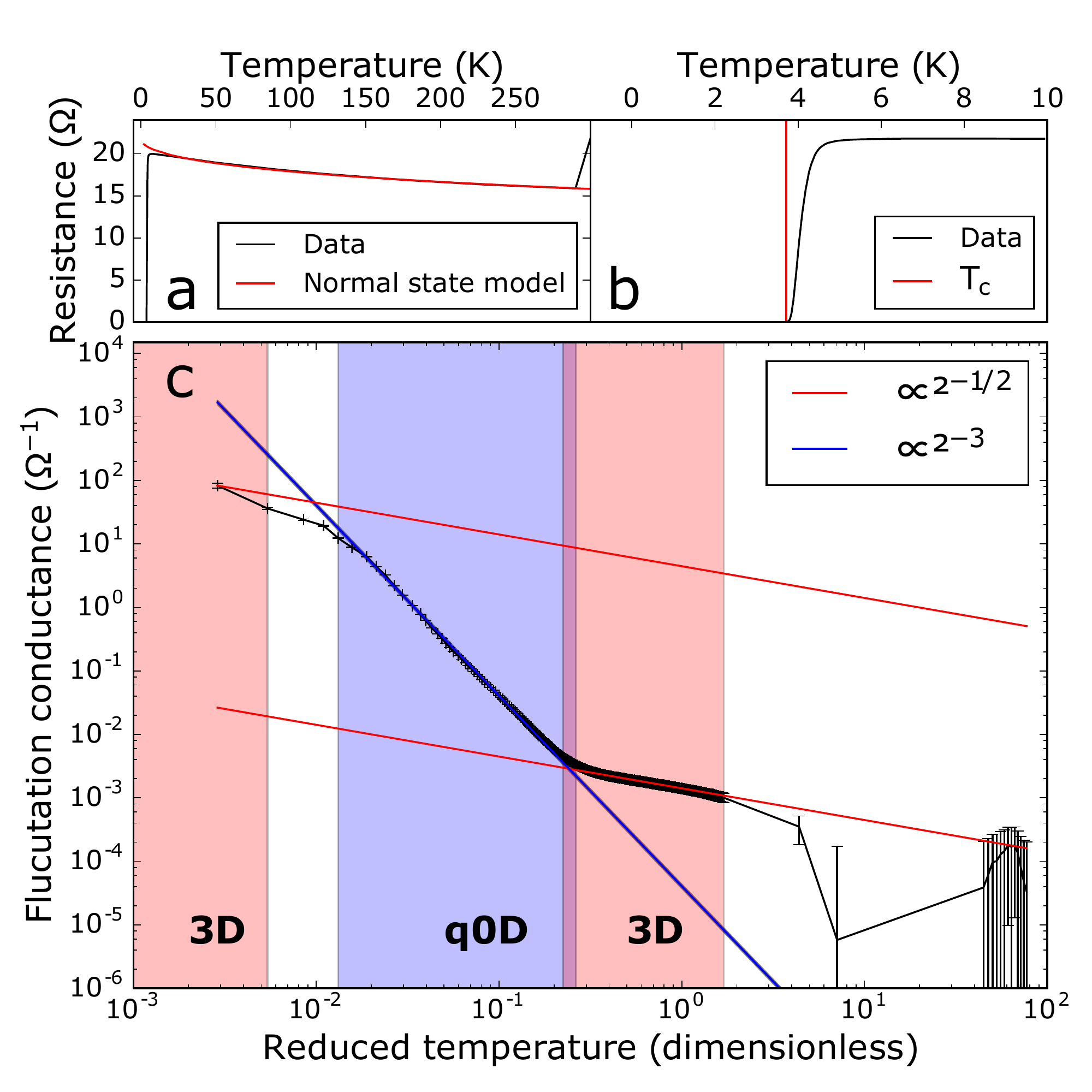}
\caption{\label{204nm}Fluctuation spectroscopy of a 204~nm thick BNCD film. (a) A fit to the high temperature region reveals a $T^{0.5}$ dependence, and allows extraction of $G_\text{ns}$. (b) $T_c$ is defined as the point at which the conductance diverges, depicted in red. (c) Fluctuation conductance, $G_\text{fl} = G -G_\text{ns}$, as a function of reduced temperature, $\epsilon = (T-T_c)/T_c$.}
\end{figure}

\begin{figure}
\includegraphics[width=8.6cm]{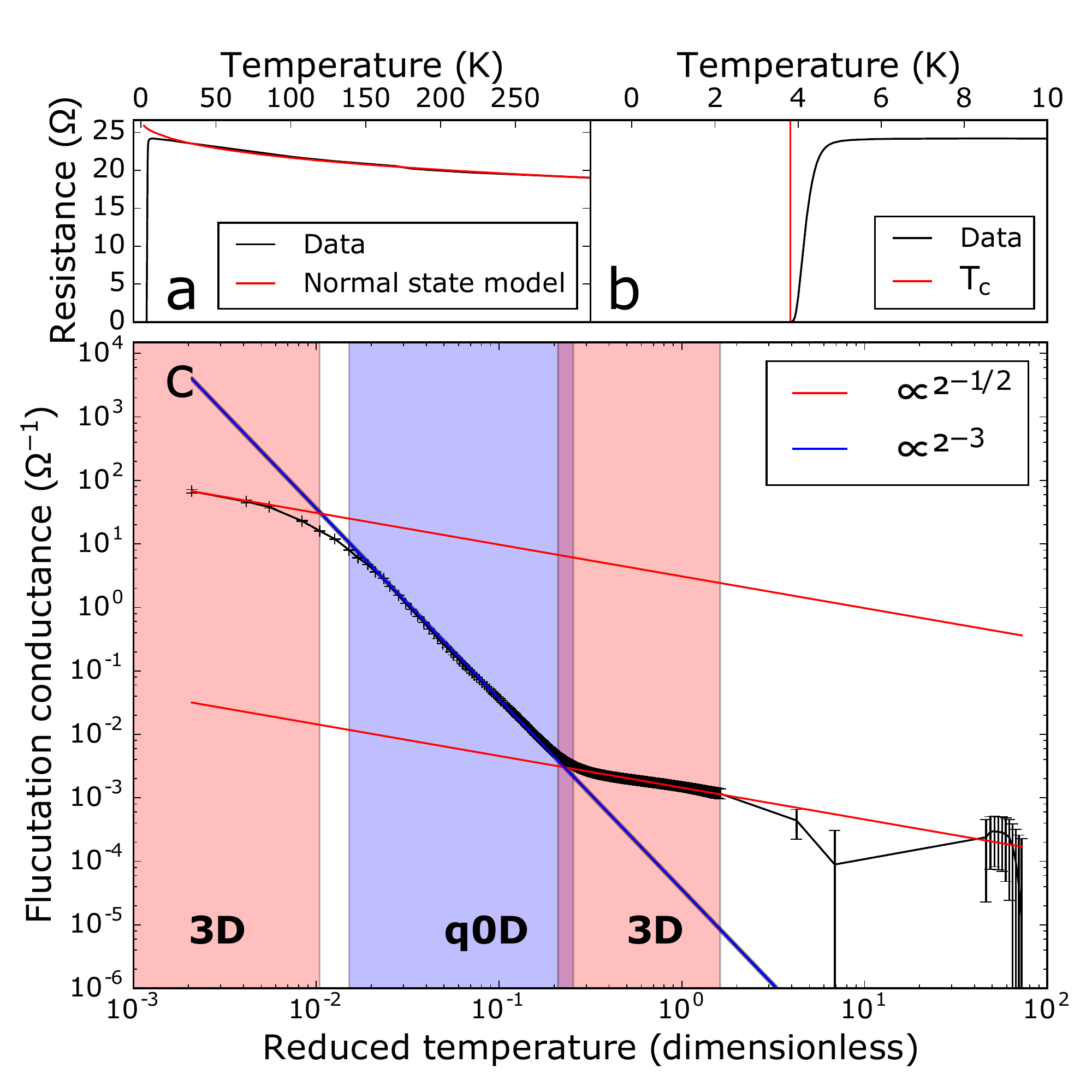}
\caption{\label{339nm}Fluctuation spectroscopy of a 339~nm thick BNCD film. (a) A fit to the high temperature region reveals a $T^{0.5}$ dependence, and allows extraction of $G_\text{ns}$. (b) $T_c$ is defined as the point at which the conductance diverges, depicted in red. (c) Fluctuation conductance, $G_\text{fl} = G -G_\text{ns}$, as a function of reduced temperature, $\epsilon = (T-T_c)/T_c$.}
\end{figure}

\begin{figure}
\includegraphics[width=8.6cm]{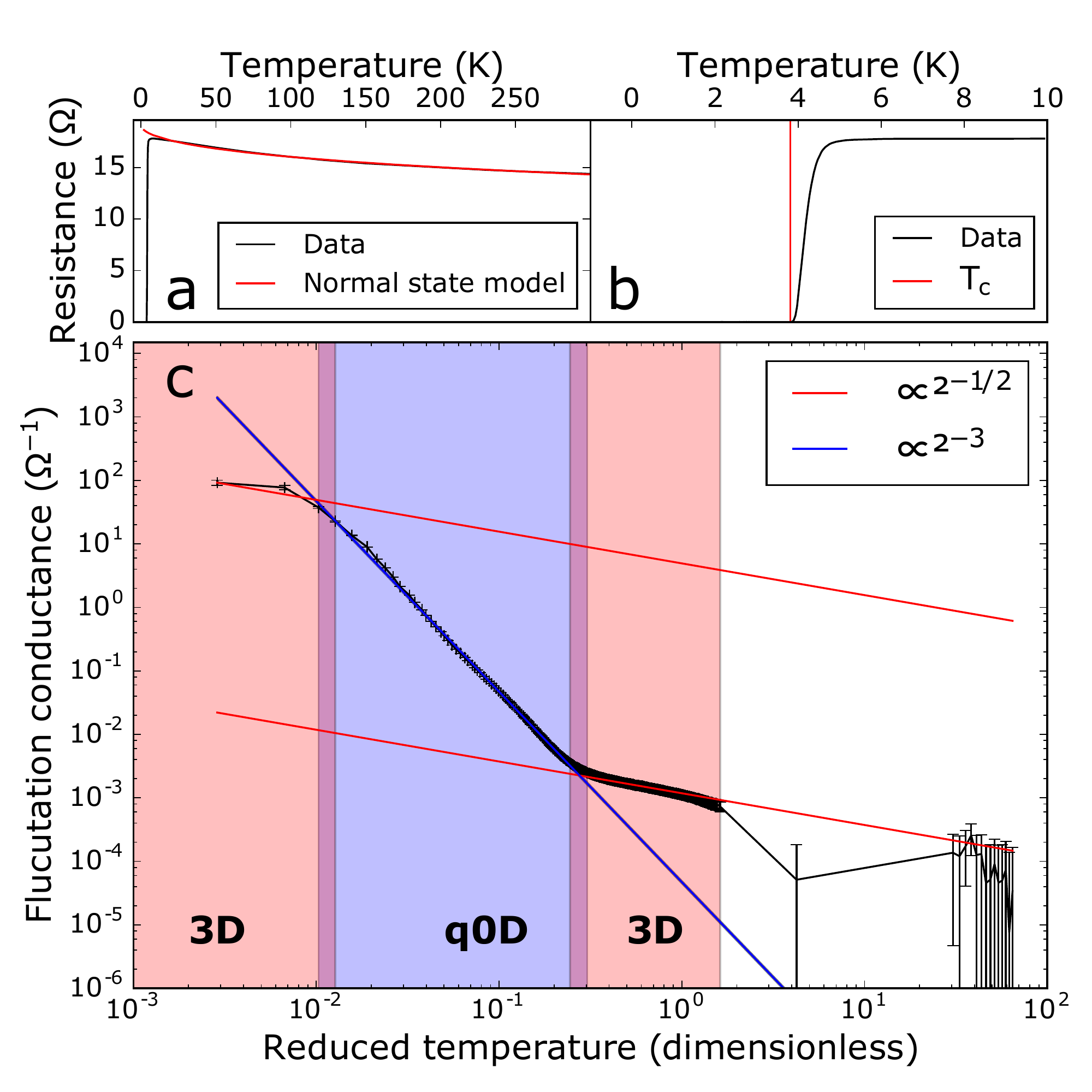}
\caption{\label{564nm}Fluctuation spectroscopy of a 564~nm thick BNCD film. (a) A fit to the high temperature region reveals a $T^{0.5}$ dependence, and allows extraction of $G_\text{ns}$. (b) $T_c$ is defined as the point at which the conductance diverges, depicted in red. (c) Fluctuation conductance, $G_\text{fl} = G -G_\text{ns}$, as a function of reduced temperature, $\epsilon = (T-T_c)/T_c$.}
\end{figure}

Fluctuation spectroscopy of all granular diamond films studied in this work are shown in Fig.~\ref{128nm}--\ref{564nm}. Each film studied showed a clear 3D -- quasi-0D -- 3D crossover, the locations of which we use to extract the tunnelling and Thouless energies.

\vspace{50pt}

\end{document}